\documentclass[aps,pra,twocolumn,floatfix,superscriptaddress,showpacs,psfig,reprint, longbilbiography]{revtex4-1}
\usepackage[english]{babel}
\usepackage[utf8]{inputenc}
\usepackage[T1]{fontenc}
\usepackage{graphicx}
\usepackage{amssymb}
\usepackage{babel}
\usepackage{amsmath}
\usepackage{amsbsy}
\usepackage{times}
\usepackage{dsfont}
\usepackage{bm}
\usepackage{float}
\usepackage{braket}
\usepackage{color,soul}
\usepackage{wasysym}
\usepackage{mathrsfs}
\usepackage{comment}
\usepackage{ulem}

\usepackage[dvipsnames,svgnames,table]{xcolor}
\usepackage{hyperref}
\usepackage{fancyhdr}
\usepackage{float}
\usepackage[english]{babel}
\usepackage[caption=false]{subfig}
\usepackage{braket}
\hypersetup{
pdfstartview={FitH},% fits the width of the page to the window
colorlinks=true,    % false: boxed links; true: colored links
linkcolor=NavyBlue, % color of internal links
citecolor=blue,%Maroon,   % color of links to bibliography
filecolor=NavyBlue, % color of file links
urlcolor=blue   % color of external links
}

\begin{document}
\title{Strain-induced phase diagram of the $S = \frac32$ Kitaev material $\rm{CrSiTe_3}$}
\author{Zongsheng Zhou}
\affiliation{School of Physical Science and Technology $\&$ Key Laboratory for Magnetism and Magnetic Materials of the MoE, Lanzhou University, Lanzhou 730000, China}
\affiliation{Lanzhou Center for Theoretical Physics and Key Laboratory of Theoretical Physics of Gansu Province, Lanzhou University, Lanzhou 730000, China.}
\author{Ken Chen}
\affiliation{School of Physical Science and Technology $\&$ Key Laboratory for Magnetism and Magnetic Materials of the MoE, Lanzhou University, Lanzhou 730000, China}
\affiliation{Lanzhou Center for Theoretical Physics and Key Laboratory of Theoretical Physics of Gansu Province, Lanzhou University, Lanzhou 730000, China.}
\author{Qiang Luo}
\affiliation{Department of Physics, University of Toronto, Toronto, Ontario M5S 1A7, Canada}
\author{Hong-Gang Luo}
\email{luohg@lzu.edu.cn}
\affiliation{School of Physical Science and Technology $\&$ Key Laboratory for Magnetism and Magnetic Materials of the MoE, Lanzhou University, Lanzhou 730000, China}
\affiliation{Lanzhou Center for Theoretical Physics and Key Laboratory of Theoretical Physics of Gansu Province, Lanzhou University, Lanzhou 730000, China.}
\affiliation{Beijing Computational Science Research Center, Beijing 100084, China}
\author{Jize Zhao}
\email{zhaojz@lzu.edu.cn}
\affiliation{School of Physical Science and Technology $\&$ Key Laboratory for Magnetism and Magnetic Materials of the MoE, Lanzhou University, Lanzhou 730000, China}
\affiliation{Lanzhou Center for Theoretical Physics and Key Laboratory of Theoretical Physics of Gansu Province, Lanzhou University, Lanzhou 730000, China.}

\begin{abstract}
The interplay among anisotropic magnetic terms, such as the bond-dependent Kitaev interactions and single-ion anisotropy,
plays a key role in stabilizing the finite-temperature ferromagnetism in the two-dimensional compound $\rm{CrSiTe_3}$.
While the Heisenberg interaction is predominant in this material, a recent work shows that it is rather sensitive to the compressive strain,
leading to a variety of phases, possibly including a sought-after Kitaev quantum spin liquid
[C. Xu, \textit{et. al.}, Phys. Rev. Lett. \textbf{124}, 087205 (2020)].
To further understand these states, we establish the quantum phase diagram of a related bond-directional spin-$3/2$ model
by the density-matrix renormalization group method.
As the Heisenberg coupling varies from ferromagnetic to antiferromagnetic,
three magnetically ordered phases, i.e., a ferromagnetic phase, a $120^\circ$ phase and an antiferromagnetic phase, appear consecutively.
All the phases are separated by first-order phase transitions, as revealed by the kinks in the ground-state energy and the jumps
in the magnetic order parameters. However, no positive evidence of the quantum spin liquid state is found and possible reasons are discussed briefly.
\end{abstract}

\maketitle
\section{Introduction}
The bond-directional interactions have emerged as a focus of research in hunting for nontrivial quantum states
ever since Kitaev proposed his famous spin-$1/2$ model on the honeycomb lattice in 2006, later dubbed Kitaev honeycomb model \cite{Kitaev2006}.
As a rare exactly solvable model which supports non-Abelian statistics and Majorana fermion excitations,
the Kitaev model is believed to be a milestone in the area of quantum spin liquids~(QSLs) \cite{Anderson1973,Balents2010,WenNPJ2019}.
However, experimental realization of such interactions is highly nontrivial due to their bond-dependent Ising nature.
Later on, Jackeli and Khaliullin provided a feasible routine to realize these interactions
by the collaboration of spin-orbit coupling and electron correlations \cite{Jackeli2009}.
Of particular interest are the $5d^5$ iridates $A\rm{_2IrO_3}$$(A\rm{=Na, Li)}$ \cite{Chaloupka2010}
and $\alpha$-$\rm{RuCl_3}$~\cite{PlumbKW2014,KimHeungSik2015,Banerjee2016},
which are recognized as effective $j = 1/2$ Mott insulators with essential Kitaev interactions.
These works soon ignited prosperous experimental and theoretical studies on such Kitaev materials~
(see \cite{WangDongetal2017,WinterNcom2018,Maksimov2020,LiZhangWang2021,HermannsM2018,Takagi2019} and references therein).

On the other hand, the non-Kitaev interactions are usually nonnegligible in candidate materials.
For example, the Heisenberg ($J$) interaction is assumable to be large in $\rm{Na_2IrO_3}$ \cite{Chaloupka2010},
while the symmetric off-diagonal $\Gamma$ term \cite{Rau2014a}, together with the distortion-induced $\Gamma^\prime$ term \cite{Rau2014b},
is essential in $\alpha$-$\rm{RuCl_3}$.
This leads to a prototypical $\rm JK\Gamma\Gamma^\prime$ model,
which serves as a versatile playground to explore exotic phases and collective phenomena.
These include a multi-node QSL in the vicinity of the Kitaev limit \cite{WangBL2019},
a gapless QSL proposed in the honeycomb $\Gamma$ model \cite{LuoZhaoetal2021,CatunYWetal2018,GohlkeWYetal2018},
and a chiral-spin ordering favored by a tiny $\Gamma'$ interaction in the dominated $\Gamma$ region \cite{LuoCSL2020}.

While the spin-$1/2$ iridates and $\alpha$-$\rm{RuCl_3}$ have been extensively studied,
recent efforts are also extended to $3d^7$~\cite{SanoRyoya2018,LiuHuimei2018,Vivanco2020,Songvilay2020}
and $f^1$~\cite{JangSeongHoon2019,JangSeongHoon2020,Motome2020B} electron states. In these materials holding large effective spins,
bond-directional Ising interactions have also been proposed~\cite{Stavropoulos2019,Stavropoulos2021}.
Among them $X\rm{_3Ni_2SbO_6}$$\left(X=\rm{Li, Na}\right)$~\cite{Zvereva2015,Kurbakov2017} are candidates of $S=1$ Kitaev materials,
while $\rm{CrXTe_3\left(X=Si,Ge\right)}$ and $\rm{CrI_3}$ are $S=3/2$ ones~\cite{XuChangsong2018,XuChangsong2020,Lee2020,ChenLebing2020,Bacaksiz2021}.
In these high-spin materials, a single-ion anisotropy (SIA) coming from the partially unquenched orbital moment of the magnetic ion should also be
taken into account \cite{XuChangsong2018,XuChangsong2020,Bacaksiz2021,Lado2017}.

Recently, density functional theory calculations on $\rm{CrSiTe_3}$ show that the Heisenberg interaction can be tuned significantly
while the other magnetic terms are nearly unchanged under compressive strain~\cite{XuChangsong2018,XuChangsong2020,Bacaksiz2021}.
Interestingly, a possible Kitaev QSL state, which is inferred from the double-peak specific heat anomaly and the associated thermal entropy plateau,
is proposed to sit between a ferromagnetic (FM) phase and an antiferromagnetic (AFM) phase \cite{XuChangsong2020}.
Their argument in mind is that the Kitaev QSL is known to display two peaks in the specific heat \cite{NasuUM2015,KogaTN2018},
which are the reminiscence of the itinerant Majorana fermions and vison excitations \cite{KnolleKCM2014}.
Nevertheless, the proposed QSL state is far from mature for the following reasons.
In the Kitaev model, intensity of the low-temperature peak in the specific heat is usually comparable to
or is even larger than the high-temperature peak \cite{NasuUM2015,KogaTN2018,LuoHuKee2021}.
However, it is reported in Ref.~[\onlinecite{XuChangsong2020}] that the low-temperature peak is rather weak in $\rm{CrSiTe_3}$.
In addition, occurrence of the double-peak specific heat does not necessarily imply a QSL ground state,
and several magnetically ordered states are demonstrated to possess such a specific heat anomaly in the  
extended Kitaev-Heisenberg model~\cite{YamajiSY2016}. These observations motivate us to revisit the QSL state in the relevant parameter region.

\begin{figure*}[ht]
\includegraphics[scale=0.44]{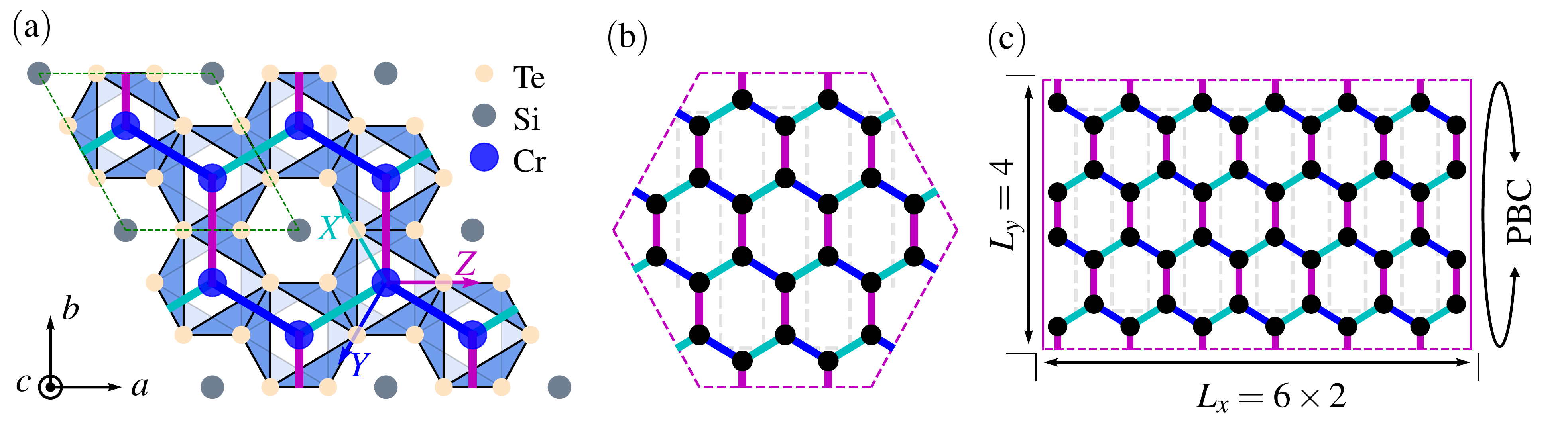}
\caption{(a) A sketch diagram of the top view of the ${\rm CrSiTe_3}$ monolayer, a unit cell is marked by the dashed parallelogram.
${\rm Cr}$ atoms locate at the centers of octahedra formed by six ${\rm Te}$ atoms,
and these ${\rm Cr}$ atoms form a two-dimensional honeycomb network.
In the $\left\{{XYZ}\right\}$ coordinate system, the three directions are marked by cyan, blue, and magenta arrows, respectively.
The exchange matrix on each color bond represented by $X, Y$ and $Z$ is given in the main text.
The orthogonal $\left\{abc\right\}$ coordinate system is given as follows. ${a}$ and ${b}$ are perpendicular and parallel to the Z bonds, respectively, ${c}$ is in the $\left[111\right]$ direction that is perpendicular to the honeycomb network.
(b) The 24-site $C_6$ cluster~($C_6$ is the symmetry of the honeycomb lattice, not the model). Black dots represent $S=\frac{3}{2}$ spins,
	three different Ising bonds are shown in different colors as in panel (a). Similarly, panel (c) plots 48-site cluster,
	$L_x$ is perpendicular to the Z bonds while $L_y$ is parallel to them. We adopt periodic
boundary condition along $L_y$ direction to roll the lattice into a cylinder, which is refered as the YC cylinder.
	In (b) and (c), the dashed line at the edges means the periodic
boundary condition while the solid line means open boundary condition.
The snake-like light gray line illustrates the one-dimensional mapping in our DMRG calculations.}
\label{Fig1}
\end{figure*}

In this work, we investigate the model proposed to describe $\rm{CrSiTe_3}$ by the density-matrix renormalization group~(DMRG) method
\cite{White1992,Peschel1999,Schollwock2005}.
As we tune the Heisenberg interaction from FM to AFM as a mimicry of the strain process,
the ground state changes from the FM phase to the AFM phase, through a wide intermediate region of $120^\circ$ phase
rather than a QSL proposed in the previous work on a 12-site cluster \cite{XuChangsong2020}.
To illuminate the discrepancy, we perform DMRG calculations on four different 12-site clusters at the proposed parameter.
Based on the results, we conjecture that the evidence of the possible QSL state may originate from
the artificial frustration induced by the mismatch between the $120^\circ$ order on small cluster size.

This paper is organized as follows.
In Sec. \uppercase\expandafter{\romannumeral2}, we briefly introduce the model and present some crucial details of the numerical calculations.
In Sec. \uppercase\expandafter{\romannumeral3} and Sec. \uppercase\expandafter{\romannumeral4},
we present the results of DMRG calculation and classical Monte Carlo simulation, respectively.
In Sec. \uppercase\expandafter{\romannumeral5} we show the DMRG results on four different 12-site clusters for comparison.
Finally, we summarize our work in Sec.\uppercase\expandafter{\romannumeral6}.

\section{Model and methods}
The crystal structure of $\rm{CrSiTe_3}$ is illustrated in Fig.~\ref{Fig1}(a).
The  $\rm{Cr^{3+}}$ ions form a two-dimensional honeycomb network, and the effective low-energy state 
of $\rm{Cr^{3+}}$ ion is described by an $S = 3/2$ spin.
Anisotropic interactions resulting from the strong spin-orbit coupling of the heavy ligand of $\rm{Cr}$, such as the Kitaev interaction,
the off-diagonal $\rm{\Gamma}$ and $\rm{\Gamma^\prime}$ interactions as well as the SIA, are essential to stabilize
the ferromagnetism at finite temperature.
A possible minimal model Hamiltonian up to nearest-neighbor interactions is given by~\cite{XuChangsong2018,XuChangsong2020,Stavropoulos2021, Bacaksiz2021}
\begin{equation}
\mathcal{H} = \sum_{\langle ij \rangle} \mathbf{S}_i^{\rm{T}}\mathcal{J}_{ij}^\alpha \mathbf{S}_j+\sum_i \mathbf{S}_i^{\rm{T}}\mathcal{A}\mathbf{S}_i.\label{Ham}
\end{equation}
The first term in Eq.~\eqref{Ham} is known as the $\rm{JK\Gamma\Gamma^\prime}$ model.
Here, $\langle{ij}\rangle$ are summed over nearest neighbors, and $\alpha \left({=X,Y,Z}\right)$ represents the bond shown in Fig.~\ref{Fig1}(a).
In the second term the sum runs over all sites.
$\mathbf{S}_i=\left(S_i^{\rm{X}}, S_i^{\rm{Y}}, S_i^{\rm{Z}}\right)^{\rm{T}}$ represent the spin operators at the site $i$.
The bond-dependent interactions and SIA matrices in the $\{XYZ\}$ coordinate system have the following form \cite{XuChangsong2020}
\begin{equation*}
\mathcal{J}^{\rm{X}}=\left(\begin{array}{ccc}
J+K & \Gamma^\prime & \Gamma^\prime\\
\Gamma^\prime & J & \Gamma\\
\Gamma^\prime & \Gamma & J
\end{array}\right), \qquad\mathcal{J}^{\rm{Y}}=\left(\begin{array}{ccc}
J & \Gamma^\prime & \Gamma\\
\Gamma^\prime & J+K & \Gamma^\prime\\
\Gamma & \Gamma^\prime & J
\end{array}\right)
\end{equation*}
\begin{equation*}
\mathcal{J}^{\rm {Z}}=\left(\begin{array}{ccc}
J & \Gamma & \Gamma^\prime\\
\Gamma & J & \Gamma^\prime\\
\Gamma^\prime & \Gamma^\prime & J+K
\end{array}\right), \qquad\mathcal{A}=\frac{A_{cc}}{3} \left(\begin{array}{ccc}
1 & 1 & 1\\
1 & 1 & 1\\
1 & 1 & 1
\end{array}\right).
\end{equation*}
Of important note is that in the crystallographic ${abc}$ coordinate system,
the SIA term could be rewritten elegantly as $A_{cc}(S_i^c)^2$.
For $S = 3/2$ Kitaev materials, the mechanism for anisotropic interactions is different from the $S = 1/2$ case~\cite{Jackeli2009,XuChangsong2018,Stavropoulos2021}.
Therefore, the Kitaev interaction can be AFM and the off-diagonal $\Gamma$ and $\Gamma^\prime$ terms could be comparable~\cite{Maksimov2020, XuChangsong2020}.
We focus on the ground phase diagram of $\rm{CrSiTe_3}$ induced by compressive strain,
and the parameters in \eqref{Ham} are referred from the supplemental material of Ref.~\cite{XuChangsong2020}.
Since $K$, $\Gamma$, $\Gamma^\prime$ and $A_{cc}$ are almost independent of the strength of the strain,
we take them as constants by setting $K=0.275$ meV, $\Gamma=0.020$ meV, $\Gamma^\prime=-0.083$ meV, $A_{cc}=0.222$ meV.
Thus, $J$, in unit of meV, is the only variable in the phase diagram.

\begin{figure}[!htbp]
\includegraphics[scale=0.4]{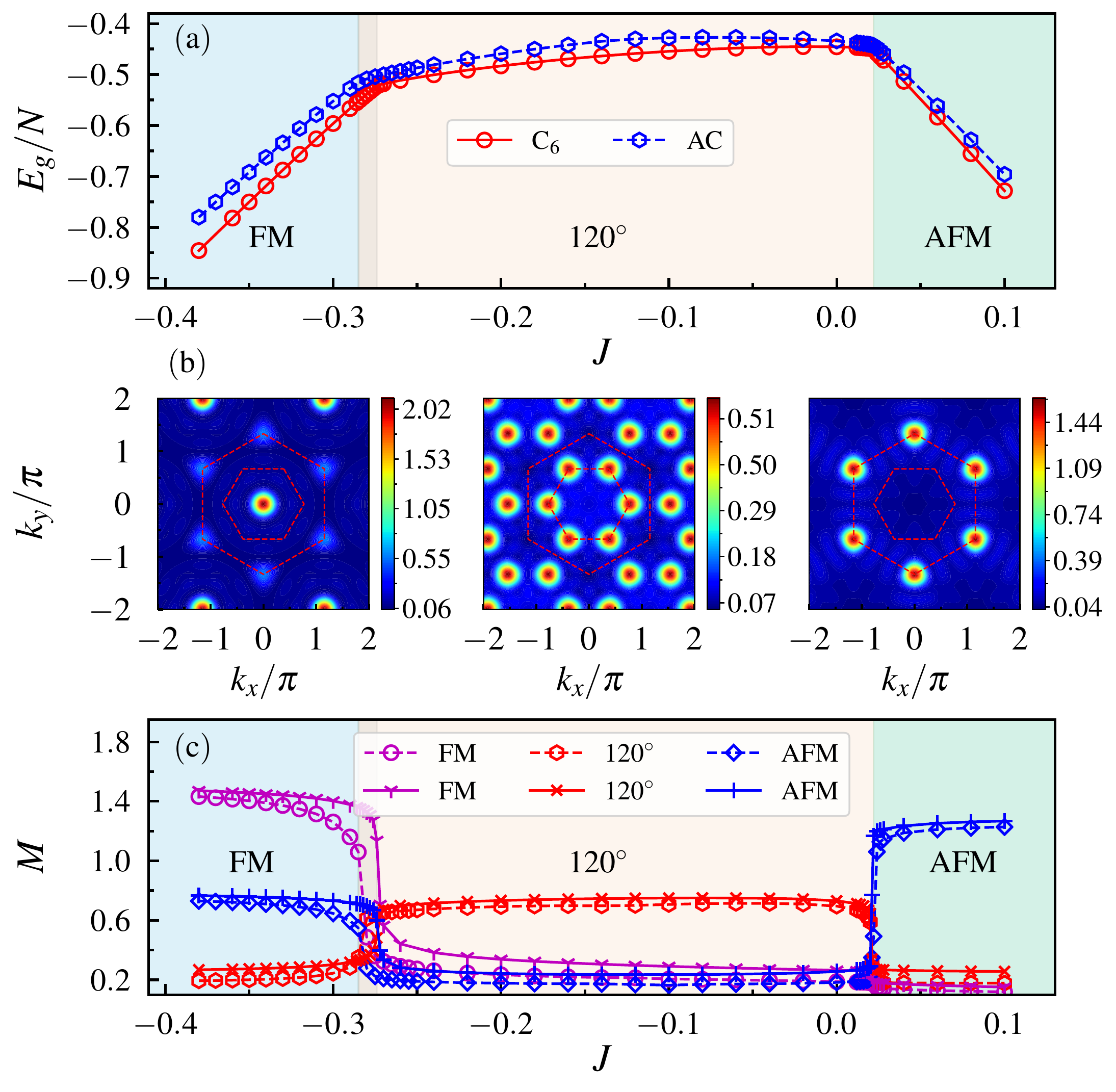}
\caption{(a) Ground-state energy per site $E_g/N$ is shown as functions of $J$ for two different clusters (see Fig.~\ref{Fig1}(b) and (c)).
In both cases, we can find three regions separated by kinks in the ground-state energy. The positions of these kinks for the two different
clusters are almost the same, which are, from left to right, $J=-0.282(3)/-0.274(2)$ and $0.022(2)$, respectively. The transition point separating the FM phase and the $120^\circ$ phase has a minor difference in these two clusters. This distinction is marked out in light gray.
(b) The typical SSSF in each phase on $C_6$ cluster is shown. (c) The order parameters $M$ (see the text) are shown as functions of $J$
	for the YC cluster and ${C_6}$ cluster, respectively.
The characteristic momenta to extract the order parameters from the SSSF are
$\mathcal{Q}_{\Gamma}$, $\mathcal{Q}_{{\rm{K}}}$, and $\mathcal{Q}_{\Gamma^\prime}$, which correspond to the FM, $120^\circ$ and AFM orders, respectively.
The positions of the discontinuities in the dominant order parameters and the kinks in the ground-state energy agree well. } \label{QPD}
\end{figure}

The phase diagram of the model~\eqref{Ham} induced by the strain is identified through the ground-state energy,
order parameters as well as static spin structure factors~(SSSFs), which are readily available in our DMRG calculation.
In the simulations, we use different clusters to confirm that our numerical results are reliable. One should be careful 
whether the clusters match the magnetic orders. As shown in Fig.~\ref{Fig1}(b) and Fig.~\ref{Fig1}(c),
24-site $C_6$ cluster with periodic boundary condition and 48-site YC cylinder ($6\times4\times2$) are mainly considered in our simulations.
In the calculations, the number of states we keep ranges from 1200 to 1600, and at least 12 sweeps are implemented
in each calculation to ensure the convergence of DMRG results. After the calculations are converged,
the utmost magnitude of truncation error in each sweep is $10^{-5}$.

\section{Quantum Phase Diagram}
Our DMRG results are summarized in Fig.~\ref{QPD}. In panel (a), we plot the ground-state energy per site $E_g/N$ as
a function of $J$ on the 24-site ${C_6}$ cluster~(Fig.~\ref{Fig1}(b)) and the 48-site YC cylinder~(Fig.~\ref{Fig1}(c)). 
Three phases are clearly detected
which are separated by kinks in the energy curve. A kink in the energy curve indicates the level crossing and thus represents a first-order phase transition.
Moreover, the positions of these kinks agree well on these two clusters. It is interesting to notice that
the exact diagonalization~(ED) on a small cluster with 12 sites~\cite{XuChangsong2020} already provides an excellent estimate of the phase boundaries.
Such a weak finite-size effect is the typical feature of first-order phase transitions.
These three phases turn out to have FM, $120^\circ$ and AFM orders.
One may notice that the energy of the $C_6$ cluster is lower than that of the YC cluster. This is because in the $C_6$ cluster PBC is used and
more bonds are connected which contribute negative energy to $E_g$.

To gain more information on these phases, we calculate the SSSF which is defined as
\begin{equation*}
  S^{\alpha\alpha}\left(\mathcal{Q}\right)  = \frac{1}{N^2}\sum_{ij}\langle S_i^\alpha S_j^\alpha \rangle e^{i\mathcal{Q}\cdot\left(\mathbf{r}_i-\mathbf{r}_j\right)}
\end{equation*}
where $N$ is the number of sites and $i$, $j$ run over all the sites. $\langle\cdots\rangle$ is the expectation in the ground state.
The order parameter at the momenta $\mathcal{Q}$ is then given by $M=\sqrt{\sum_{\alpha}{{S}^{\alpha\alpha}(\mathcal{Q})}}$.
In the magnetic ordered phase, it is finite at the characteristic momentum in the thermodynamic limit.
Therefore, we can tell the magnetic orders at the characteristic momenta determined by the peaks in the SSSF.
In panel (b), we show the typical contour plots of the SSSF $\sum_\alpha S^{\alpha\alpha}(\mathcal{Q})$ in the three different phases.
To keep the symmetry of the Brillouin zone, we use the data from the $C_6$ cluster. The data from the cylinder give the same conclusion
except that the symmetry of the pattern is lower.
Hence, the characteristic momenta in these three phases are
$\mathcal{Q}_{\Gamma}=\left(0, 0\right)$,
$\mathcal{Q}_{{\rm{K}}}=\left({2\pi}/{3}, {2\pi}/(3\sqrt{3})\right)$,
and $\mathcal{Q}_{\Gamma^\prime}=\left({2\pi}/{3}, {2\sqrt{3}\pi}/{3}\right)$, respectively,
as well as others associated by the symmetry transformation of the Brilouin zone.
In panel (c), we show the order parameters at the characteristic momenta as a function of $J$.
The order parameters at the characteristic momenta in other phases are also shown for comparison.
This figure tells us that the three phases from left to right are an FM phase, a $120^\circ$ phase and an AFM phase.
At all the phase transition points, the discontinuity in the order parameters is clearly visible,
indicating a first-order transition. This is consistent with the conclusions drawn from the ground-state energy.

\begin{figure}[!tbbp]
\includegraphics[scale=0.46]{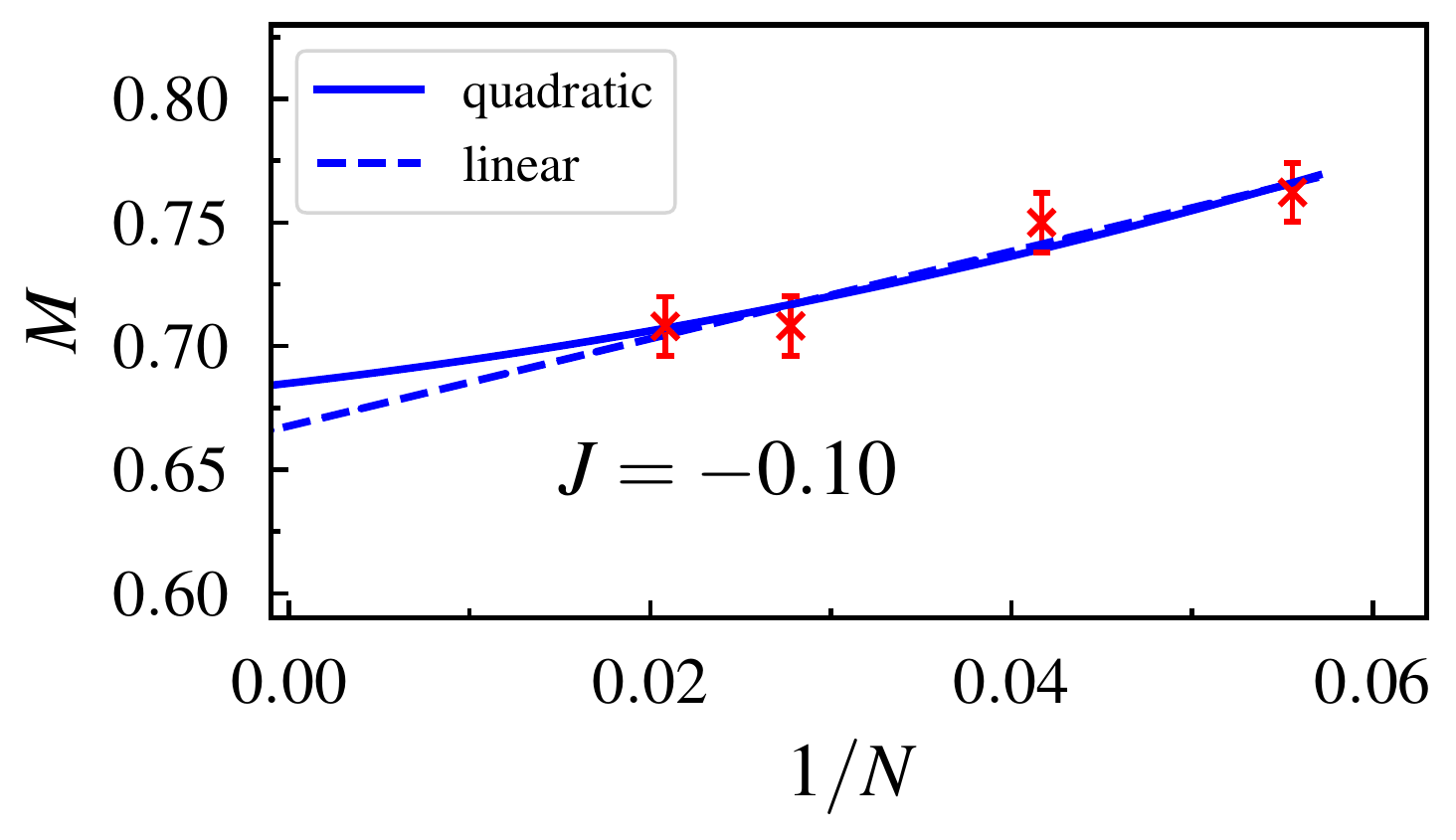}
\caption{Extrapolation of the order parameter at $\mathcal{Q}_{\rm{K}}$ in $120^\circ$ phase.
  The geometries used are rhombus clusters of 18 and 24 sites, and XC cylinder of $6\times3\times2$ and YC cylinder of $6\times4\times2$.
  The dashed line is the linear fitting and the solid line is the quadratic fitting, and the magnetization is estimated as $M=0.67(2)$.
  }\label{FSS}
\end{figure}

To check the tendency of the $120^\circ$ magnetic ordering against the system size,
we calculate the order parameter of the $120^\circ$ phase on four clusters.
Here, the Heisenberg interaction $J$ is taken as $-0.10$, which is close to the strength of strain $-2.34\%$~\cite{XuChangsong2020}.
As can be seen from Fig.~\ref{FSS},
the order parameter is rather robust and the magnetization is around one-half of the spin value $3/2$, albeit with a decreasing trend as the system size increases.
The linear and quadratic fittings suggest that the magnetization is approximately 0.67(2) in the thermodynamic limit.
It is worthing to note that the magnetization is obvious larger than the well-recognized $120^\circ$ order in the spin-$1/2$
triangular-lattice Heisenberg antiferromagnet \cite{QianLi2020},
demonstrating the robustness of the $120^\circ$ order in the present honeycomb model.

\section{Classical phase diagrams}
Large spins usually indicate weak quantum fluctuation. An $S = 3/2$ spin is considered to be near the boundary between the classical
and the quantum world~\cite{Songvilay2021}. On the other hand, quantum fluctuation can be enhanced by anisotropic magnetic terms,
So to what extent does an $S = 3/2$ spin model behave classically in frustrated systems?
We will try to answer this question by a direct comparison between the quantum phase diagram and its classical counterpart.
For this purpose, in this section we will establish the classical phase diagram by the parallel-tempering Monte Carlo method \cite{Hukushima1996,Mitsutake2000,JanssenAV2016} .
The simulations are performed on a 12$\times$12$\times$2 rhombic cluster. The spin configurations are updated by the heat bath algorithm.

\begin{figure}[!htbp]
\includegraphics[scale=0.36]{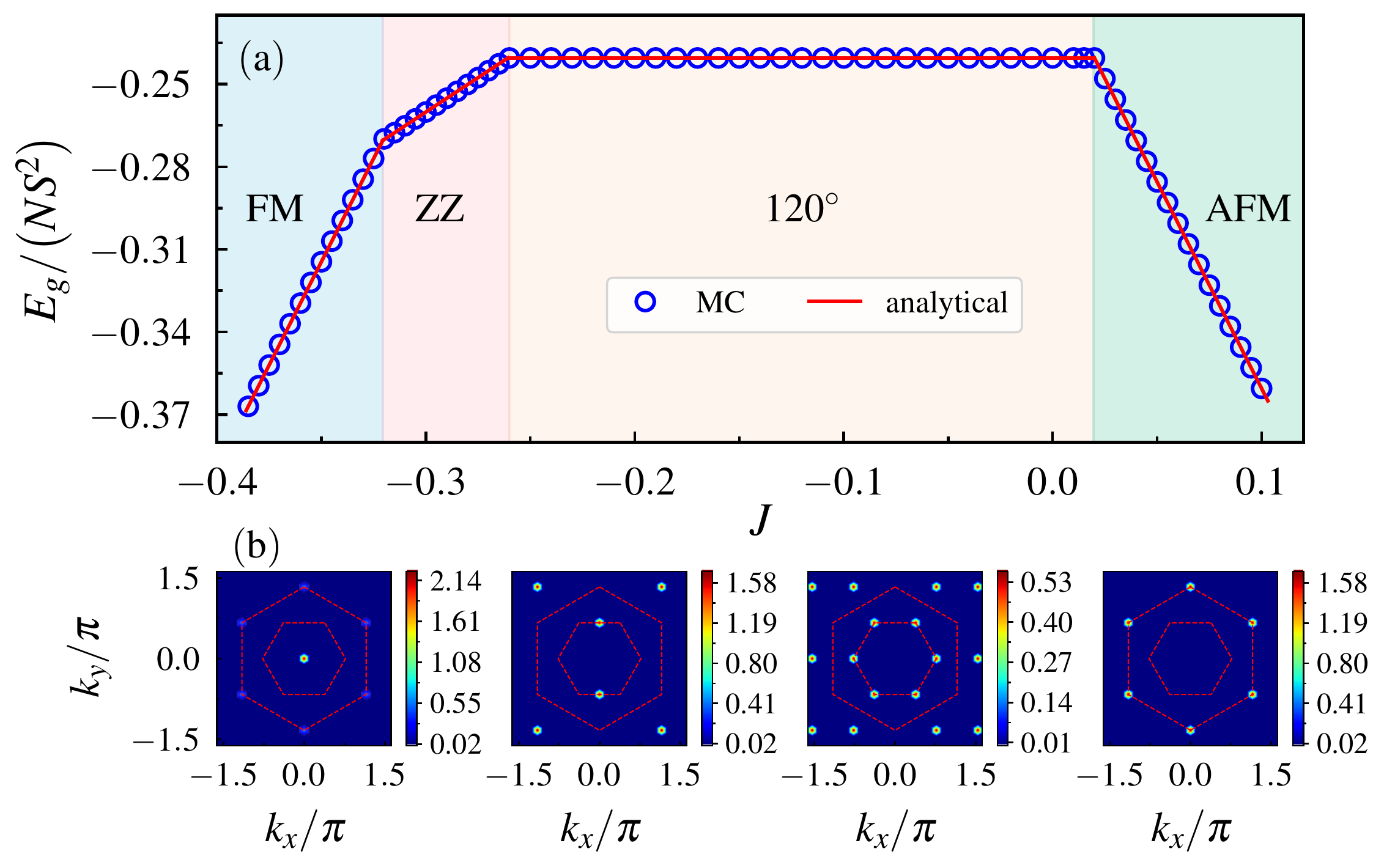}
\caption{(a) The blue open circles are the ground energy obtained by the Monte Carlo. There are four phases in the phase diagram,
i.e., an FM phase, a zigzag phase, a $120^\circ$ phase and an AFM phase from left to right. The transition points are
$J\approx{-0.32063}, -0.26074, 0.020$. The red lines are analytical results of the ground-state energy.
(b) The SSSFs in the corresponding phases.}
\label{CF}
\end{figure}

In the classical limit, each spin is regarded as an $O\left(3\right)$ vector in real space. We use the Monte Carlo method to find the ground state.
The corresponding ground-state energy and spin configurations are then readily available.
In Fig.~\ref{CF} (a), we plot the classical ground-state energy per spin $E_g/\left(NS^2\right)$ as a function of $J$.
Interestingly, it is easy to see that the ground-state energy is a piece-wise
function of $J$ and four phases can be determined in the phase diagram.
The spin configurations in each phase directly tell us the characteristic features,
which are an FM phase, a zigzag (ZZ) phase, a $120^\circ$ phase and an AFM phase, respectively, from left to right.
Moreover, the expressions of the classical energy can be given analytically
\begin{eqnarray*}
\begin{split}
&E_{\rm{FM}}\!=\!\frac{S^2}{2}\!\left[3J\!+\!K\!-\!\Gamma\!-\!2\Gamma^\prime\! +\!P(\Gamma,\! \Gamma^\prime,\! A_{cc})\Theta\!\left(P(\Gamma,\! \Gamma^\prime,\! A_{cc})\right)\right],\\
&E_{\rm{ZZ}}=\frac{S^2}{4}\left(2J\!-\!\Gamma\! +\!2\Gamma^\prime\!+\!2A_{cc}\right) + S^2 Q\left(K, \Gamma, \Gamma^\prime, A_{cc}\right), \\
&E_{120^{\circ}}=\frac{S^2}{2}\left(-K-2\Gamma+2\Gamma^\prime\right),\\
&E_{\rm{AFM}}=\frac{S^2}{2}\left(-3J-K+\Gamma+2\Gamma^\prime\right),
\end{split}\label{En}
\end{eqnarray*}
where $\Theta\left(x\right)$ is a step function and $P\left(\Gamma, \Gamma^\prime, A_{cc}\right)$ and $Q\left(K, \Gamma, \Gamma^\prime, A_{cc}\right)$ have the following forms
\begin{align*}
	P  = & 3\Gamma + 6\Gamma^\prime + 2A_{cc},\\
	Q  = &-\frac{1}{12}\sqrt{32\left(K\!-\!\Gamma\!+\!\Gamma^\prime\right)^2\!+\!\left(6A_{cc}\!+\!2K\!+\!7\Gamma\!+\!2\Gamma^\prime\right)^2}.
\end{align*}
Here, the occurrence of the step function $\Theta(x)$ in the FM phase is because that its classical moment direction
could either lie in the $ab$ plane (when $x > 0$) or point along the $c$ direction (when $x < 0$).
In Fig.~\ref{CF}(a), we also plot these analytical results for comparison, which are marked by red lines.
The overlap of these data demonstrates the reliability of our numerical results. Just as in the quantum case,
the phases can also be identified by the SSSF, which are plotted in Fig.~\ref{CF}(b). The SSSF in the FM phase,
the $120^\circ$ phase and the AFM phase show a close resemblance to the quantum cases in Fig.~\ref{QPD}(b).
In particular, the heights of characteristic peaks are close in their value, indicating that the Eq.~(\ref{Ham})
in the quantum case is close to its classical limit.
%We want to mention that in the classical situation the spin configurations will not be the superposition of all the degenerated states,
%and some weights of the degenerate states are missing in the ZZ phase.

\begin{figure}[hbt]
\includegraphics[scale=0.5]{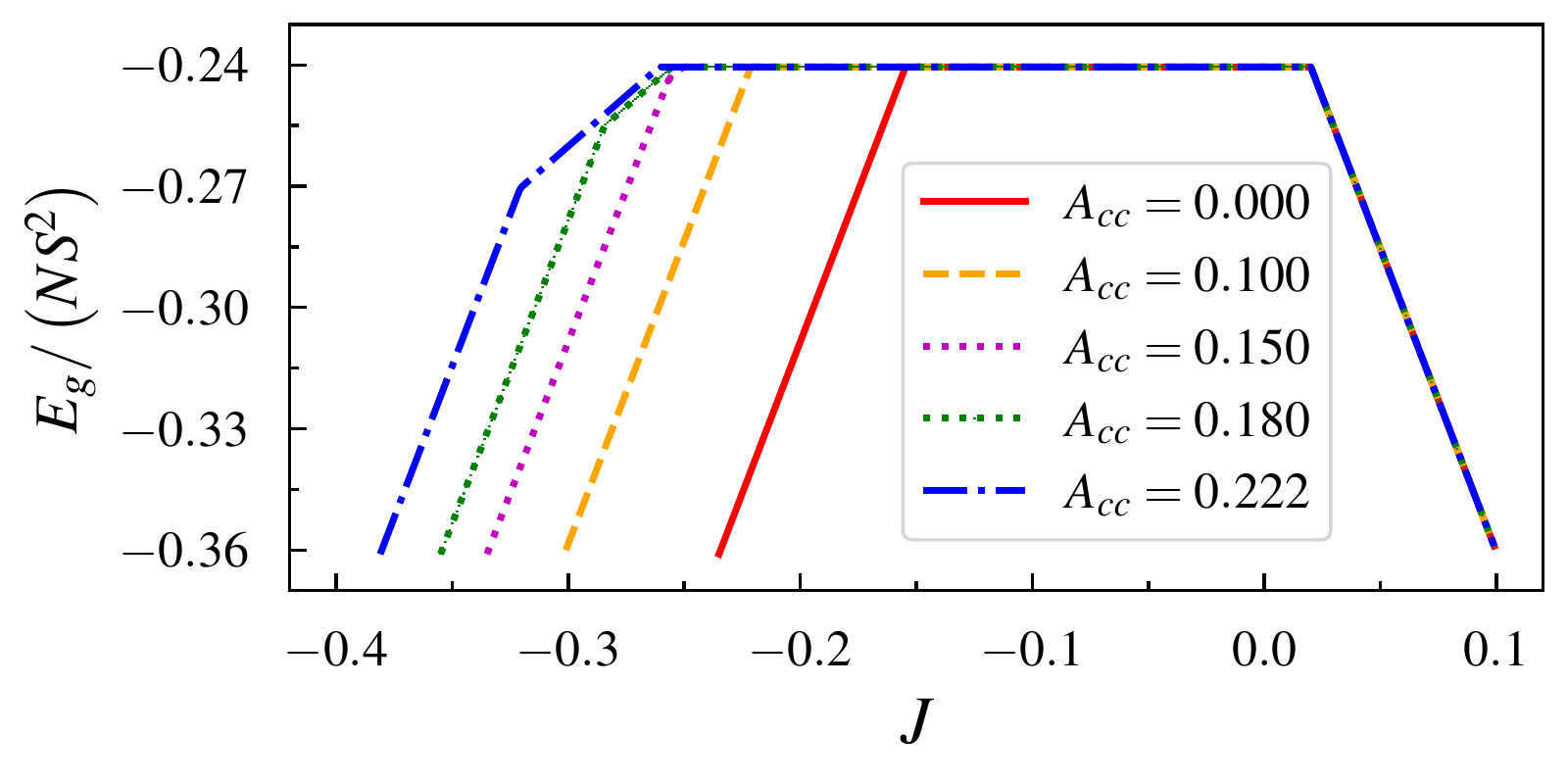}
\caption{Classical ground-state energy is plotted as a function of $J$ for $A_{cc} = 0.000, 0.100, 0.150, 0.180$ and $0.222$.
We can see that when $A_{cc}\lesssim 0.150$, zigzag phase disappears.}
\label{CE}
\end{figure}

From the Fig.~\ref{QPD} and Fig.~\ref{CF}, we can see that the transition point from the $120^\circ$ phase to the AFM phase
agrees well in both the quantum and classical cases, which are $J=0.022(2)$ and $J=0.020$, respectively.
However, an unexpected zigzag phase emerges in the classical phase diagram, which sits between the FM phase and $120^\circ$ phase.
We have checked that when $A_{cc}$ is smaller than $0.150$, the zigzag phase disappears as shown in Fig.~\ref{CE}.
This is not surprising because the classical spin corresponds to $S\rightarrow\infty$ and in the $S=3/2$ case
the quantum fluctuation can not be completely neglected and thus is expected to shift the phase boundary.

\section{Finite-size effect in 12-site clusters}

Generally speaking, the QSL is known as a nonmagnetic state with fractionalized excitations,
which could be manifested by topological entanglement entropy \cite{KtvPre2006,LvnWen2006}, continuum excitation spectrum,
and thermodynamic signatures \cite{DoPark2017}.
In the Kitaev QSL, the spins fractionalize into localized and itinerant Majorana fermions, resulting in a double-peak specific heat anomaly \cite{NasuUM2015,KogaTN2018}.
Nevertheless, the double-peak structure of specific heat found in the previous work on a 12-site cluster
can not simply be interpreted as a signature of spin fractionalization \cite{XuChangsong2020}.
Instead, it may originate from the artificial frustration due to the incompatible cluster with respect to the underlying magnetic ordering.

\begin{figure}[hbt]
\includegraphics[scale=0.45]{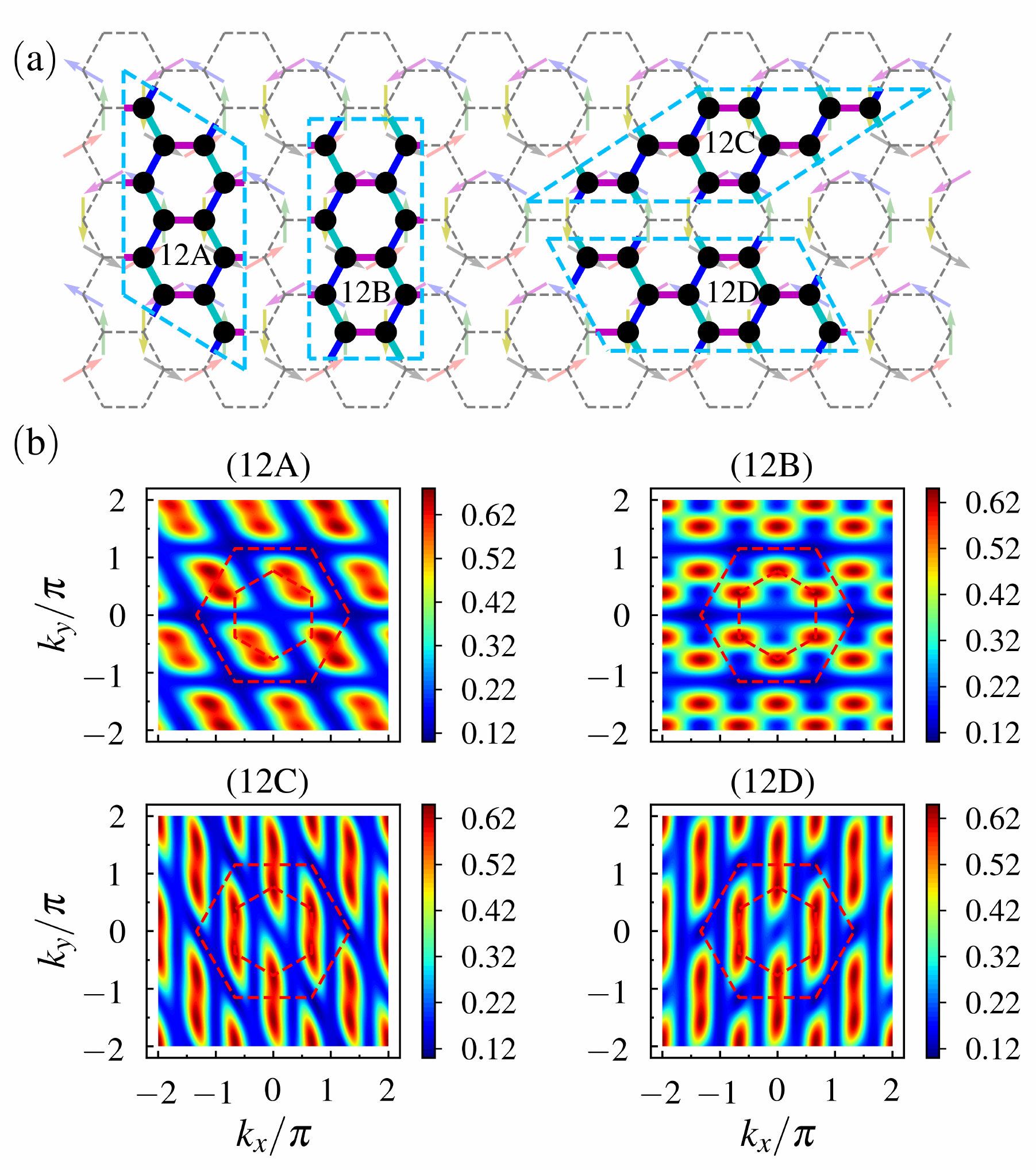}
\caption{(a) Four different 12-site clusters with full periodic boundary conditions.
  The background shows the classical spin configuration of the six-sublattice $120^\circ$ ordering.
  The 12A cluster have a energy of -4.835103988, while the last three clusters (12B, 12C, and 12D) have a same but lower total energy of -5.535684265.
  (b) SSSF on the four clusters, which are 12A (top left), 12B (top right), 12C (bottom left) and 12D (bottom right).}
\label{Fig:12Site}
\end{figure}

We have illustrated in Sec. III that the intermediate region is a $120^\circ$ magnetically state rather than the QSL.
For comparison, we fix $J$ = $-0.10$ again where the ground state locates deep in the $120^\circ$ phase.
To be compatible with the six-sublattice $120^\circ$ ordering, the number of lattice sites in a cluster should be a multiple of six.
Even though such condition is satisfied, the cluster shape should be chosen carefully so as to match the magnetic order.
As stated in Ref.~[\onlinecite{XuChangsong2020}], the 12A cluster in Fig.~\ref{Fig:12Site}(a) is likely used in the specific heat calculation.
Unfortunately, this geometry does not coordinate the $120^\circ$ order in the vertical boundary,
which inevitably introduces the artificial frustration and ruins the magnetic ordering significantly in the small cluster.
To confirm this, we have performed the DMRG calculations on the 12A cluster.
The total ground-state energy is -4.835103988, and the corresponding SSSF is shown in the first panel of Fig.~\ref{Fig:12Site}(b).
There is not a well-defined peak in the Brillouin zone, in accord with the proposed QSL inferred from the specific heat.

By contrast, a slightly modification of the geometry will remove the mismatch of the spins at boundaries,
see 12B cluster in Fig.~\ref{Fig:12Site}(a). Its SSSF in the second panel of Fig.~\ref{Fig:12Site}(b) exhibits a well-defined peak
at $\mathcal{Q}_{{\rm{K}}}$ point, in line with the desirable $120^\circ$ order. 
We have also considered two other 12-site clusters (termed 12C and 12D) which match the $120^\circ$ ordering, 
and their SSSFs with peaks at $\mathcal{Q}_{{\rm{K}}}$ point are shown in the last row in Fig.~\ref{Fig:12Site}(b).
Notably, All the clusters that coordinate the $120^\circ$ spin textures (i.e., 12B, 12C, and 12D) share a same total energy of -5.535684265,
which is considerably lower than that of the 12A cluster.
This demonstrates again that the QSL based on the 12A cluster is not the true ground state.
We want to stress that even in the 12B, 12C and 12D clusters which have the same ground-state energy the difference in their SSSFs is clearly visible. 
In particular, the SSSFs in 12C and 12D are rather diffusive in the momentum space in comparison with that of 12A case, 
demonstrating that the results on large clusters are necessary to reveal the underlying phase.

\section{summary and outlook}
In this work, we study an $S = 3/2$ $\rm JK\Gamma\Gamma^\prime$+SIA model proposed to describe $\rm{CrSiTe_3}$ by using the DMRG method.
Upon increasing the strain, the Heisenberg~($J$) interaction varies significantly while the remaining ones almost keep unchanged.
we thus map out the $J$-dependent phase diagram as a mimicry of the strain process.
When $J$ varies from negative to positive, three magnetically ordered phases are found, including an FM phase, a $120^\circ$ phase and an AFM phase.
All the phase transitions are of the first order, characterized by the kinks in the ground-state energy and discontinuities in the order parameters.
Experientially, the quantum fluctuation is expected to be strongly suppressed in the models with a large spin of $S = 3/2$,
and the physics should bear resemblance to its classical counterpart\cite{Lee2020}.
We have strengthened this conclusion by performing the classical Monte Carlo simulations on the same model,
with the exception that a zigzag phase is found between the FM phase and the $120^\circ$ phase.
The zigzag phase is sensitive to the SIA, and its territory shrinks gradually and it disappears eventually as SIA decreases.

It was argued in a previous work~\cite{XuChangsong2020} that there is a possibly QSL,
which is inferred from the double-peak structure of specific heat and the related plateau of thermal entropy on a 12-site cluster.
Curiously, the proposed QSL region happens to coincide with the $120^\circ$ phase identified in our work.
In fact, the double-peak specific heat anomaly solely could not legitimate the Kitaev QSL.
To begin with, the low-temperature peak found in Ref.~[\onlinecite{XuChangsong2020}] is rather weak when compared to the 
high-temperature analogy, which is in contrast to the current wisdom that the magnitudes of the two peaks are comparable 
in the Kitaev QSL \cite{NasuUM2015,KogaTN2018}. In addition, it is unclear if such a low-temperature peak will 
still persist with the increasing of the system size. Even though the double-peak structure indeed exists in the ground state,
it is still not enough to authenticate a QSL as magnetically ordered states could also exhibit such an abnormal phenomenon \cite{YamajiSY2016}.
In this regard, it is imperative to check if the ground state is disordered or not.
Our large-scale DMRG calculation, however, suggests that the ground state is a magnetically ordered $120^\circ$ state with a large order parameter.

Although our work implies that QSL is unfavorable in $\rm{CrSiTe_3}$, three magnetic ordered phases are found under the strain.
These ordered states may persist to finite temperature because the Mermin-Wagner theorem does not forbid such orders in the model (\ref{Ham}).
In this sense, this material offers a playground to study these magnetic orders and phase transitions in low dimensions\cite{LiXuetal2021}.
Recent experiment and \textit{ab initio} calculations support $\rm{Cr}$-based ferromagnetic van der Waals materials are $S = 3/2$ Kitaev
materials~\cite{XuChangsong2018, XuChangsong2020,Lee2020,Bacaksiz2021}, the model here just takes the nearest-neighbor interactions into account,
the other possible interactions are not considered. Due to the complexity of real material, the neutron spectra of $\rm{CrI_3}$ can
be fitted by different models~\cite{ChenLebing2020}. In $\rm{CrSiTe_3}$, we can not exclude other possible terms such as the next-nearest 
neighbors in the Hamiltonian.
Moreover, the model Hamiltonian~\eqref{Ham} and corresponding parameters remain to be verified by experiments.
We expect our results can be verified in future experiments and be helpful to understand the magnetic behavior in two-dimensional ferromagnetism in high-spin Kitaev materials.

%%%%%%%%%%%%%%%%%%%%%%%%%%%%%%%%%%%%%%%%%%%%%%%%%%%%%%%%%%%%%%%%%%%%%%%%%%%%%%
\begin{acknowledgments}
We thank Y. Qi, F. Zheng, C. Xu and L. Bellaiche for helpful discussion.
This work is supported by the National Natural Science Foundation of China (Grant No. 11874188, 12047501, 11774300).
The computational resource is partly supported by Supercomputing Center of Lanzhou University.
\end{acknowledgments}

%%%%%%%%%%%%%%%%%%%%%%%%%%%%%%%%%%%%%%%%%%%%%%%%%%%%%%%%%%%%%%%%%%%%%%%%%%%%%%%

\nocite{*}
%\bibliographystyle{apsrev4-1_title}
%\bibliography{Reference}% Produces the bibliography via BibTeX.

\clearpage
%\begin{widetext}
%\clearpage
%\centerline{ {\Large \bf Supplementary Materials For $S=\frac{3}{2}$ Kitaev Magnets} }
%\appendix
%\title{Supplementary Materials:
%Phase diagram of the spin-1/2 Kitaev-Gamma chain
%and emergent SU(2) symmetry
%}
%\maketitle
%\tableofcontents
%\newpage
%%%%%%%%%%%%%%%%%%%

%%%%%%%%%%%%%%%%%%%
%\begin{thebibliography}{10}
%\end{thebibliography}

%\end{widetext}
\end{document}